\documentclass[aps,prl,reprint,showpacs,showkeys,groupedaddress]{revtex4-1}

\usepackage{graphicx}
\usepackage{bm}
\usepackage{dcolumn}
\usepackage{color}

\newcommand{\bea}{\begin{equation}}
\newcommand{\eea}{\end{equation}}
\newcommand{\ber}{\begin{eqnarray}}
\newcommand{\eer}{\end{eqnarray}}

\begin{document}


\title{Quantum potential induced emergence of massive scalar fields in the analogue gravity model of a Bose-Einstein condensate}



\author{Supratik Sarkar}
\author{A Bhattacharyay}
\email[]{a.bhattacharyay@iiserpune.ac.in}




\affiliation{Indian Institute of Science Education and research, Pune, India}


\date{\today}

\begin{abstract}
We show here a general approach to include the quantum potential term in the emergent gravity model of Bose-Einstein condensate by using multiple scales. Our main result shows the emergence of a massive scalar modulating field at larger length scales as a result of Lorentz symmetry breaking at the length scales comparable to the healing length. We also propose that, the nonlocal interactions induced tuning of healing length can be exploited experimentally to observe the systematics of small and large scale coupling as emerges in our present analysis.    
 
\end{abstract}


\maketitle

{\it Introduction} -
It is possible to create large curvatures of effective space-time as seen by sonic excitations in some condensed matter systems at a very low temperature. Bose-Einstein condensate (BEC) is one of the prominent candidates of all these systems \cite{barc1}. This fact opens up the scope of experimenting on various aspects of quantum fields on curved space-time, e.g., possibility of observing Hawking radiation, cosmological particle production etc. Unruh's seminal work \cite{unruh} practically opened up this field of research which is actively pursued over last couple of decades and a host of theoretical proposals are around \cite{jacob,barc2}. Particularly in the context of emergent gravity in BEC, Parentani and coworkers have proposed a number of novel ideas based on density correlations, studying the hydrodynamics over several length scales, surface gravity independent temperature etc \cite{par1,par2,par3}. Novel use of density-density correlations inside and outside the horizon has also been exploited by Balbinot {\it et al} \cite{bal}. 
\par
BEC is a superfluid quantum phase of matter which is considered to be the most probable condensed matter candidate to create analogue gravitational scenarios at a very low (nano Kelvin) temperature \cite{zoll1,zoll2}. The small amplitude collective excitations of a uniform density moving phase of BEC obeys the quantum hydrodynamics which, ignoring the  quantum potential term \cite{pita}, can be cast into the equation of a massless free scalar field on a Lorentzian manifold. There has been efforts to regularize the dynamics taking into account the  quantum potential term \cite{fleu}.{ { Visser et al. have shown the emergence of a massive Klein-Gordon equation considering a two component BEC where a laser induced transition between the two components is exploited \cite{viss}. Liberati et al. proposed a weak $U$(1) symmetry breaking of the analogue BEC model by the introduction of an extra quadratic term in the Hamiltonian to make the scalar field massive \cite{lib}. Considering the flow in a Laval nozzle, Cuyubamba has shown the emergence of a massive scalar field in the context of analogue gravity arguing for the possibility of observation of quasi-normal ringing of the massive scalar field within the laboratory setup \cite{cuyu}.}}  
\par
The healing length of a BEC, in the standard condensed matter context, is considered to be $\lambda_C/\sqrt{2}$ where $\lambda_C$ is the effective Compton wavelength of a particle where the velocity of light is replaced by the velocity of sound. Below this length scale, there happens a Lorentz breakdown in the analogue picture and dispersion becomes important. At these length scales, sound waves start seeing the structure of the matter and the corresponding dispersion relation is known for many condensed matter systems. One takes advantage of this fact of knowing the dispersion relation and tries to address the analogue trans-planckian problem in this regime \cite{unr,un1,sch,brou,corl}. The basic idea of most of such works is to understand the robustness of the Hawking radiation (Planckian spectrum) in the presence of Lorentz-breaking dispersions.
\par
Our main result in this letter is to show that as a consequence of  quantum potential induced Lorentz symmetry breaking of the massless scalar field at smallest length scales, there emerges a massive scalar field at larger length scales on a spacetime of different geometry. This is an important result, within the scope of analogue systems, in view of the fact that it indicates the analogue trans-planckian effects at larger scale dynamics. Our main result for a quantum fluid is general in the sense that, it does not need to take into account anything special about the condensate. We simply consider here a single component free condensate without any special confinement or symmetry breaking. As our simple derivation reveals, the main result is always there on the consideration of the  quantum potential term in BEC as a perturbation. By considering nonlocal interactions in a BEC at its minimal level of s-wave scattering, we additionally show that, the small length scale at which the Lorentz symmetry breaking happens can possibly be tuned to a good extent. This tuning of the Lorentz breaking length scale can be very important for the experimental observation of the intrinsic coupling between small and large scales, however, our main result is independent of this tuning. 
\par      
{\it Minimal GP model for nonlocal s-wave scattering} -
A nonuniform BEC is characterized by the mean field Gross-Pitaevskii (GP) equation of the form
\ber\nonumber
 && i\hbar\frac{\partial}{\partial t}\psi({\bf r},t) = \left ( -\frac{\hbar^2\bigtriangledown^2}{2m} + V_{ext}({\bf r},t) \right )\psi({\bf r},t)\\ &+& \left (\int {d{\bf r^\prime}\psi^*({\bf r^\prime},t)V({\bf r^{\prime}-{\bf r}})\psi({\bf r^\prime},t)} \right )\psi({\bf r},t),
\eer
where $\psi$ is the order parameter and $|\psi|^2=n(r,t)$ is the density of the condensate. In the above equation, $m$ is the mass of a boson, $\hbar$ is the Planck constant, $V_{ext}$ is an external potential and $V(\bf r -r^\prime)$ is the interaction potential. At the limit, the s-wave scattering length $a$ of the system being much smaller than the average separation between particles $n^{-1/3}$, one makes a delta function approximation to the range of $V(\bf r -r^\prime)$ and writes the local GP equation in the form
\bea
i\hbar\frac{\partial}{\partial t}\psi({\bf r},t) = \left ( -\frac{\hbar^2\bigtriangledown^2}{2m} + V_{ext}({\bf r},t) + g|\psi({\bf r},t)|^2  \right )\psi({\bf r},t).
\eea
In the above expression, $g=4\pi\hbar^2 a/m$ is the strength of the s-wave scattering considered at the lowest order Born approximation. 
\par
Due to the possibility of increasing the s-wave scattering length practically from $-\infty$ to $\infty$ near a Feshbach resonance, which has already been experimentally achieved \cite{corn}, we can take nonlocal s-wave scattering into account to write the GP equation to the leading order approximation in a Taylor expansion of the order parameter in the following form
\ber\nonumber
i\hbar\frac{\partial}{\partial t}\psi(x,t) &=& \left ( -\frac{\hbar^2}{2m}\frac{\partial^2}{\partial x^2} + V_{ext} + g|\psi(x,t)|^2\right )\psi(x,t)\\ &+& \frac{1}{6}a^2g\psi(x,t)\frac{\partial^2}{\partial x^2}|\psi(x,t)|^2 .
\eer
{The above expression has been written in 1D cartesian coordinate for the sake of simplicity. A change of coordinate or symmetry might change the small numerical pre-factor of the last term without affecting the dependence on scattering length $a$ whose critical value would be determined by the density $n$ which is very large number ($\sim 10^{14}$). Therefore, this numerical pre-factor is not that important and can also be absorbed in the effective flat s-wave scattering potential $V_{eff}$ where $g = \int{d{\textbf r}V_{eff}}$.} In Eq.3 we are merely considering the lowest order s-wave scattering as is taken into account in writing Eq.2 with the exception that we are removing the approximation of the $\delta$-correlated particle interactions. Here, we are considering the interaction range to be the same as the s-wave scattering length $a$ which is a good approximation for the lowest energy scattering. The last term represents the correction to the local GP equation (Eq.2) at its minimal level as one considers the non-locality of the symmetric interactions.
\par
Let us note a few features of Eq.3 in comparison with the local GP equation (Eq.2). The continuity equation is preserved for our model as well giving a conservation of mass as that for local GP equation. The uniform (so-called) ground state solution $\psi_0=\sqrt{n}e^{-i\mu t/\hbar}$ of the local GP equation is also a solution of a free condensate of Eq.3. The condition of dynamical stability of this so-called ground state will remain the same as the local GP dynamics.{{ Our model is also derivable from an exact free energy functional \cite{pend}.}}
\par
Consider the small amplitude excitations to the ground state $\psi_0$ in order to have a look at the dispersion relation. To that end, let us perturb the ground state as $\psi=\psi_0+[\sum_j{u_j(x)e^{-i\omega_j t}+v_j^*(x)e^{i\omega_j t}}]e^{-i\mu t/\hbar}$ where $\mu=gn$ is the chemical potential of the system. The linearized dynamics of the small amplitudes emerges as
\ber\nonumber
\hbar\omega_j u_j &=& gnu_j + (\frac{a^2gn}{6}-\frac{\hbar^2}{2m})u_j^{\prime\prime} + gnv_j + \frac{a^2gn}{6}v_j^{\prime\prime} \\\nonumber
-\hbar\omega_j v_j &=& gnv_j + (\frac{a^2gn}{6}-\frac{\hbar^2}{2m})v_j^{\prime\prime}+ gnu_j + \frac{a^2gn}{6}u_j^{\prime\prime}\\,
\eer
with a dispersion relation
\bea
\hbar^2\omega^2=\frac{\hbar^2k^2gn}{m}+\left (\frac{\hbar^4}{4m^2} - \frac{\hbar^2a^2gn}{6m}  \right )k^4
\eea 
where the double prime indicates double spatial derivative. {Creation of roton-minimum in the dipolar BEC with {\it long-range} anisotropic interactions is a theoretical proposal over a long time \cite{wils1,wils2}. Our simple analysis partly captures the role of the long range of the interactions (keeping it isotropic) in creating a roton minimum like dispersion here. In the present context, we need not vary the range of the interaction so much that it bends the dispersion curve down beyond a finite $k$, rather, we would be needing the tuning of the s-wave scattering length to the extent that it straightens the curve to a good extent by making the quartic term in the dispersion relation small.} 
\par
From this dispersion relation, one can find out the healing length based on usual balance between the kinetic and potential energy of the system \cite{sup} as
\bea
\xi = \xi_0\left ( \frac{1}{2} -\frac{a^2}{6\xi_0^2} \right )^{1/2},
\eea
where $\xi_0=\hbar/\sqrt{2mgn}=1/\sqrt{8\pi an}$ is the healing length of the system when the correction term due to nonlocal interactions is not taken into account. Note that, the expression for the velocity of sound does not change from $c=\sqrt{gn/m}$, but, the healing length can vary much more rapidly near $a = n^{-1/3}$ than the $1/\sqrt{a}$ scaling as $a$ increases. Excitations which are at a smaller length scales than the healing length are considered particle like whereas the ones at a larger length scales are sound waves (characterized by a linear dispersion relation) which actually see the effective curved space-time. It's obvious that, now, by tuning $a$ one can make $\xi$ become vanishingly small and we would be using this fact in our perturbative treatment to derive the field equations {in order to have the effects of the possible tuning on our results manifestly present}. Note that, this shrinking of $\xi$ can be viewed as an increase of effective mass towards infinity which explains the shrinking of the effective Compton wavelength of the system.
\par
{\it Emergent gravity} -
Considering a general single particle state of the BEC given by $\psi=\sqrt{n}e^{i\theta({\bf r},t)/\hbar}$, one adds perturbations to the density and the phase as $n \rightarrow n + {n}_1$ and $\theta \rightarrow \theta + {\theta}_1$ to get the linearized dynamics in $n_1$ and $\theta_1$ as
\ber
\frac{\partial {n}_1}{\partial t}+\frac{1}{m}\bigtriangledown . ({n}_1\bigtriangledown \theta + n\bigtriangledown{\theta}_1)=0,\\
\frac{\partial {\theta}_1}{\partial t} + \frac{\bigtriangledown\theta . \bigtriangledown{\theta}_1}{m} + g^\prime{n}_1 - \frac{\hbar^2}{2m}D_2{n}_1=0.
\eer
In the above equation we have replaced $g$ by $g^\prime$ to avoid a clash of notations later. The operator $D_2$ is given by 
\ber\nonumber
D_2{n}_1 &=& -\frac{{n}_1}{2}n^{-3/2}\bigtriangledown^2{n^{1/2}}+\frac{n^{-1/2}}{2}\bigtriangledown^2{(n^{-1/2}{n}_1)}\\ &-& \frac{g^\prime
ma^2}{3\hbar^2}\bigtriangledown^2{{n}_1},
\eer 
where the last term is due to our nonlocal correction to the local GP equation. Considering a uniform background density of the system ($n$ is a constant) one can write this operator as
\bea
D_2 = \frac{2m}{\hbar^2}\left( \frac{\hbar^2}{4mn}-\frac{a^2g^\prime}{6}\right ) \bigtriangledown^2. 
\eea
If one does not take into account the nonlocal interactions at the least as has been considered by us here, the second term on the r.h.s. in the expression of $D_2$ in Eq.10 would be missing.
Now, one expresses $n_1$ in terms of $\theta_1$ and puts that in the continuity Eq.7 to derive the equation for the phase field as
\bea
\partial_{\mu}f^{\mu \nu}\partial_{\nu}\theta_1 =0.
\eea
The matrix elements $f^{\mu \nu}$ are
\ber\nonumber
f^{00} &=& -(g^\prime -\frac{\hbar^2}{2m}D_2)^{-1}\\\nonumber
f^{0j} &=& -(g^\prime -\frac{\hbar^2}{2m}D_2)^{-1}v^j\\\nonumber
f^{i0} &=& -v^i(g^\prime -\frac{\hbar^2}{2m}D_2)^{-1}\\\nonumber
f^{ij} &=& \frac{n\delta^{ij}}{m}-v^i(g^\prime -\frac{\hbar^2}{2m}D_2)^{-1}v^j.
\eer
\par
In the standard practice, within the Thomas-Fermi limit, one neglects the terms containing $D_2$ considering large scale variation of the density (equivalently phase) fluctuations. Note that, the coefficient of the $D_2$ is not actually a smaller term then $g^\prime$ because $g^\prime = 4\pi \hbar^2 a/m$. In the absence of nonlocal interactions, consider the common factor in all the $f^{\mu\nu}$ in its inverse form as 
\bea\nonumber
g^\prime - \frac{\hbar^2}{4mn}\bigtriangledown^2 = \frac{\hbar^2}{2mn}(\frac{1}{\xi_0^2}-\frac{1}{2}\bigtriangledown^2).
\eea 
It's obvious from the above equation that, to be able to drop the second term in the expression compared to the first, the length scale of variation must be much larger than $\xi_0$. Starting from the defining inequality of the local GP dynamics $a << n^{-1/3}$, one can show that $\xi_0 >> n^{-1/3}$. So, the length scale under consideration is effectively much larger than the average inter-particle 
separations because so is the $\xi_0$. On the basis of this approximation, one casts Eq.11 in the standard covariant form
\bea
\frac{1}{\sqrt{-g}}\partial_\mu(\sqrt{-g}g^{\mu \nu}\partial_\nu)\theta_1 = 0,
\eea
where $f^{\mu \nu} = \sqrt{-g}g^{\mu \nu}$.
\par
Considering the tunability of $\xi$, at a small $\xi$, to the leading order in $\xi^2$, the elements of $f^{\mu \nu}$ can be written as 
\ber\nonumber
f^{00} &=&-\frac{1}{g^\prime}(1 + \xi^2\bigtriangledown^2)\\\nonumber 
f^{0j} &=& -\frac{1}{g^\prime}(1 + \xi^2\bigtriangledown^2)v^j\\\nonumber
f^{i0} &=& -v^i\frac{1}{g^\prime}(1 + \xi^2\bigtriangledown^2)\\\nonumber
f^{ij} &=& \frac{n\delta^{ij}}{m}-v^i\frac{1}{g^\prime}(1 + \xi^2\bigtriangledown^2)v^j.
\eer 
This controlled expansion helps us recover Eq.13 as it is as the $O(1)$ dynamics which would break down at a length scale comparable to the healing length. Obviously, the order $\xi^2$ ($=\xi_0^2\epsilon^2$, where $\epsilon = (\frac{1}{2} -\frac{a^2}{6\xi_0^2})^{1/2}$ is the small parameter) term will act as a source term to the ensuing large scale dynamics once we separate the dynamics on independent multiple scales considering $\partial_\mu \rightarrow \partial_\mu + \epsilon\partial_\mu^\prime$. Here, the primed scale is $\epsilon^{-1}$ times larger than the scale at which one gets the Klein-Gordon equation at the intermediate scales. So, by bringing $\xi$ to zero one can push this effect of amplitude modulation to infinitely large length scales or can practically get rid of it.
\par
Let us go by this choice of scales as suggested by the dynamics itself with the field expressed in a product form as $\theta_1^\prime({R})\theta_1({r})$ where the $\theta_1^\prime({R})$ varies at larger length scales (where $R$ and $r$ are four-vectors over large and small length scales respectively), we get the equation for the $\theta_1^\prime({R})$ at $O(\epsilon)$ and at $O(\epsilon^2)$ as
\ber
&&(\partial_\mu f^{\mu \nu}\theta_1({r}))\partial_\nu^\prime\theta_1^\prime({R}) + \partial_\mu^\prime\theta_1^\prime({R})(f^{\mu \nu}\partial_\nu\theta_1({r}))  =0,\\ 
&&\partial_\mu^\prime f^{\mu\nu}\partial_\nu ^\prime\theta_1^\prime({R}) + \frac{\partial_\mu F^{\mu \nu}\partial_\nu \theta_1({r})}{\theta_1({r})}\theta_1^\prime({R}) = 0.
\eer 
Eq.15, in its structure, is the Klein-Gordon equation for a massive scalar field where the sign of the mass term which is a function of smaller length scales (hence constant) is important and is context dependent. Eq.14 is the constraint that has to be obeyed. When Eq.14 holds based on the structure of the underlying velocity field at the intermediate length scales, the massive scalar field is free. However, when Eq.14 constrains $\theta_1^\prime(R)$, the massive scalar field should reside on a surface and is not completely free. Here, $F^{\mu \nu}$ should properly include the extra $\xi_0^2\bigtriangledown^2$ factor in each of its elements. The $F^{\mu \nu}$ is different from the $f^{\mu \nu}$ because of the spatial dependence of the background velocity on intermediate scales. 
\par
By a controlled consideration of the  quantum potential term in the hydrodynamics of BEC, which could be experimentally achievable, we see the possible emergence of mass based on how close the small scale dynamics approaches to the analogue Compton wavelength of the system. Eq.13 has negligible correction if the length scales of the free scalar field is large enough compared to $\xi$. The correction becomes large at small length scales, but, shows up in the large length scale dynamics through an intrinsic coupling between the small and large scales. Note that, in the present calculations, we have actually considered the underlying velocity field to be varying at $O(1)$ length scales to keep things simple. This would mean that the spacetime of the scalar field with the possible mass term is flat. If one does take the velocity field to depend on larger scales as well, one must introduce multiple scales at an earlier stage where the spatial variation of the phase of the background state is taken into account and that could change the whole dynamics appreciably. 
\par
Let us try to understand why the nonlocal interactions enable us tune the effective Compton wavelength to a very small value as is revealed by the present analysis. The increase of the s-wave scattering length up to the average separation between particles practically renders the system structureless by making the gaps between the particles vanish. The fluid becomes a uniform continuous object because of the consideration of the uniform density of the condensate. Since, the healing length hides the microscopic structures at which a quantum theory of gravity would be relevant, in the absence of such structures the classical gravity would obviously reign which is manifest in the validity of Eq.13 over all length scales as $\xi$ vanishes.
\par
The question that arises now is - is it possible to make $\xi$ actually vanish within the scope of our approximations? The answer is no. In taking the nonlocal interactions in its simplest form, when we had truncated the Taylor expansion of the order parameter $\psi$, we had done that on the basis of the consideration that $\psi$ is slowly varying over space such that all higher derivatives of $\psi$ than the second one are negligible. We had done such an approximation, obviously, by noting the presence of the second derivative in the kinetic term and one must keep corrections up to this order. Thus, for very small length scale variations, higher order correction terms would be important. However, if one sticks to the s-wave scattering, due to the spherical symmetry, the odd order derivative terms would be all zero. So, the next correction would be consisting of a forth order derivative of $\psi$ with an opposite sign to the existing correction term. This term, when relevant at small length scales, would modify the $\xi$ and might prevent one from making $\xi$ actually vanish. Nevertheless, the possibility of tuning the nonlocal interactions can create the opportunity to make $\xi$ small and systematically probe the analogue system at a much smaller length scale than what is achievable without nonlocal interactions. In spite of this very small scale complications, the present model is enormously revealing in showing the possible emergence of massive scalar fields from the  quantum potential term {through a kind of ultraviolet-infrared coupling of scales}.
\par
{Its important to note that, the correction to the Eq.13 that has been considered here showing the emergence of mass at larger length scales is not dependent on the ability to tune the healing length under the presence of nonlocal interactions. The appearance of this correction is a general result which would be always there even if we go by taking $\xi_0$ to be the healing length and the length scales of the free massless scalar field is comparable to $\xi_0$. In this case, at the first order expansion of the term $(g^\prime - \frac{\hbar^2}{4mn}\bigtriangledown^2)^{-1}$, the coefficient of the $\bigtriangledown^2$ term will come proportional to $\xi_0^2$. The expansion would be truncated here at $O(\xi_0^2)$ considering $\xi_0$ small which is always better than throwing out the term altogether and the same qualitative picture would emerge except for the fact that to make $\xi_0 \to 0$ one has to make $a \to \infty$ at a given density.} 
\par
The phenomenon of emergent gravity generally breaks down at length scales comparable to the healing length. The effects of small scale excitations would be felt at larger length scales by the emergence of a massive scalar field on a somewhat different spacetime is our main result. An intrinsic connection between dynamics at smallest Lorentz-breaking scales and larger scales is established which can provide considerable insights through the detailed computation of specific cases using this framework. Moreover, we have used nonlocal interactions in BEC and tuning of the healing length by Feshbach resonance to reflect on possible experimentally realizable changing length scales.
 
\par{\it Acknowledgement} -
AB acknowledges very useful discussions with Suneeta Vardarajan.


\begin{thebibliography}{99}
\noindent\bibitem {barc1} C. Barcel\'o, S. Liberati, and M. Visser, Int. J. Mod. Phys. A, {\bf 18}, 21, 3735 (2003).
\noindent\bibitem {unruh} W.G. Unruh, Phys. Rev. Lett., {\bf 46}, 1351 (1981).
\noindent\bibitem {jacob} T.A. Jacobson, Phys. Rev. D, {\bf 44}, 1731 (1991).
\noindent\bibitem {barc2} C. Barcel\'o, S. Liberati, and M. Visser, arXiv:gr-qc/0505065v3 (2011).
\noindent\bibitem {par1} J. Macher and R. Parentani, Phys. Rev. A, {\bf 80}, 043601 (2009).
\noindent\bibitem {par2} S. Finazzi and R. Parentani, Phys. Rev. D, {\bf 83}, 084010 (2011).
\noindent\bibitem {par3} S. Finazzi and R. Parentani, Phys. Rev. D, {\bf 85}, 124027 (2012).
\noindent\bibitem {bal} R. Balbinot, A. Fabbri, S. Fagnocchi, A. Ricati and I Carusotto, Phys. Rev. A {\bf 78}, 021603 (2008).
\noindent\bibitem {zoll1} L. J. Garay, J. R. Anglin, J. I. Cirac, and P. Zoller, Phys. Rev. Lett., {\bf 85}, 22, 4643 (2000).
\noindent\bibitem {zoll2} L. J. Garay, J. R. Anglin, J. I. Cirac, and P. Zoller, Phys. Rev. A, {\bf 63}, 023611 (2001).
\noindent\bibitem {pita}  L. Pitaevskii and S. Stringari, Bose-Einstein Condensation, (Oxford Science Publications, 2003).
\noindent\bibitem {fleu} V. Fleurov and R. Schilling, Phys. rev. A, {\bf 85}, 045602 (2012).
\noindent\bibitem {viss} M. Visser and S. Weinfurtner, Phys. Rev. D, {\bf 72}, 044020 (2005).
\noindent\bibitem {lib} S. Libareti, F. Girelli and L. Sindoni, arXiv:0909.3834 (2009).
\noindent\bibitem {cuyu} M. A. Cuyubamba, Class. Quantum Grav., {\bf 30}, 195005 (2013).
\noindent\bibitem {unr} W.G. Unruh, Phys. Rev. D {\bf 51}, 2827 (1995).
\noindent\bibitem {un1} W.G. Unruh and R. Sch\"utzhold, Phys. Rev. D {\bf 71}, 024028 (2005).
\noindent\bibitem {sch} R. Sch\"utzhold and W.G. Unruh, Phys. Rev. D {\bf 78}, 041504(R) (2008).
\noindent\bibitem {brou} R. Brout, S. Massar, R. Parentani and Ph. Spindel, Phys. Rev. D {\bf 52}, 4559 (1995).
\noindent\bibitem {corl} S. Corley and T. Jacobson, Phys. Rev. D {\bf 54}, 1568 (1996).
\noindent\bibitem {corn} S. L. Cornish, N. R. Claussen, J. L. Roberts, E. A. Cornell, and C. E. Wieman, Phys. Rev. Lett., {bf 85}, 9, 1795 (2000).
\noindent\bibitem {sup} Supratik Sarkar and A. Bhattacharyay, J. Phys. A: Math. Theor. {\bf 47}, 092002 (2014).
\noindent\bibitem {pend} A. Pendse and A. Bhattacharyay, arXiv:1407.2039 (2014).
\noindent\bibitem {wils1} R. M. Wilson, S. Ronen and J. N. Bohn, Phys. Rev. Lett. {\bf 100}, 245302 (2008).
\noindent\bibitem {wils2} R. M. Wilson, S. Ronen and J. N. Bohn, Phys. Rev. Lett. {\bf 104}, 094501 (2010). 

 



\end{thebibliography}

\end{document}